# Ethylene glycol-mediated synthesis of nanoporous anatase $TiO_2$ rods and rutile $TiO_2$ self-assembly chrysanthemums

Quanjun Li, Bingbing Liu,[*] Yingai Li, Ran Liu, Xianglin Li, Shidan Yu, Dedi Liu, Peng Wang, Bo Zou, Tian Cui, and Guangtian Zou

State Key Laboratory of Superhard Materials, Jilin University, Changchun 130012, China

**Abstract:** Nanoporous anatase $TiO_2$ rods and rutile $TiO_2$ chrysanthemums were successfully synthesized via a simple ethylene glycol-mediated synthesis route. Their morphologies, phase compositions and components were characterized by SEM, TEM, Raman and IR, respectively. The results show that a self-assembly growth takes place in the calcination under vacuum, which makes the titanium glycolate rods transform into rutile $TiO_2$/C chrysanthemums rather than anatase $TiO_2$ rods. It also indicates that the carbon plays an important role in the phase transition process which promotes the phase transition to rutile $TiO_2$ at a lower temperature (400 $^{o}$C). It provides a new approach to prepare nanoporous rutile $TiO_2$ nanomaterials under low temperature.

## 1. Introduction

$TiO_2$ materials have attracted considerable attention recently due to their unique physicochemical properties and applications in the fields of photocatalysts, gas sensors, photovoltaics, and lithium ion batteries [1-6]. It is not surprising in view of the importance of $TiO_2$ materials that a wide variety of approaches for the synthesis of $TiO_2$ have been reported, and it remains a particularly active research field. Various morphologies of $TiO_2$ materials have been synthesized by many approaches. Such as, nanoparticles were generally obtained from sol-gel method [7]. Nanowires, nanotubes and nanoribbons were widely prepared by a hydrothermal process [8, 9]. Nanowires films were synthesized by thermal evaporation, thermal deposition and anodic oxidative hydrolysis [10-12]. Recently, ethylene glycol (EG) has been widely used in the so-called polyol synthesis of metal nanoparticles due to its strong reducing powder and relatively high boiling point (~197 $^{o}$C) [13-15]. Jiang and co-workers have synthesized metal oxide (including $TiO_2$, $SnO_2$, $In_2O_3$, and $PbO_2$) nanowires by a similar method [13]. More recently, submicrometer-sized $TiO_2$ nanorods were synthesized with EG and tetra-isopropyl orthotitanate by a simple microwave irradiation method [15]. According to previous research, synthesis of various morphologies of metal oxides is possible by using a suitable metal salt in EG. However, there is no report on the synthesis of nanoporous rutile $TiO_2$ chrysanthemum which consists of numerous nanoparticles with porous structure and is different from those previous reports of flowerlike $TiO_2$ nanomaterials with solid structure [16-19]. In this study, we report the synthesis of titanium glycolate rods, nanoporous anatase $TiO_2$ rods, and nanoporous rutile $TiO_2$ chrysanthemums by the ethylene glycol-mediated synthesis process and posttreatment.

**2. Experimental**

Titanium glycolate rods were synthesized by heating a solution of titanium alkoxide in EG at 170 $^{o}$C for 2 h. In a typical synthesis, 0.1 mL titanium butoxide was added to a 50 mL flask that contained 10 mL EG. The solution was stirred for 30 min, and then heated to 170 $^{o}$C in oven without stirring for 2 h. After cooling down to room temperature, the white flocculate was harvested using centrifugation, followed by washing with deionized water and ethanol several times to remove excess EG from the sample. The precipitate was finally dried in oven at 80 °C for 8 h in air and used for further characterization. TiO$_2$ rods were obtained by calcined the titanium glycolate rods at 400 °C for 4 h in air. TiO$_2$ chrysanthemums were obtained by calcined the titanium glycolate rods at 400 °C for 4 h in vacuum and then in air. The morphologies of the samples were observed by scanning electron microscopy (SEM). Transmission electron microscopy (TEM) images were obtained on a HITACHI H-8100 microscope using an acceleration voltage of 200 KeV. Raman spectra were excited by radiation of 514.5 nm from a Renishaw inVia Raman spectrometer. IR spectra were acquired under ambient conditions using an infrared spectrometer (NICOLET AVATAR 370 DTGS).

## 3. Results and discussion

Fig. 1 shows SEM images of the titanium glycolate rods, TiO$_2$ rods, self-assembly TiO$_2$/C chrysanthemums, and pure TiO$_2$ chrysanthemums. Fig. 1(a) reveals the presence of numerous titanium glycolate rods with lengths ranging from several micrometers to several tens of micrometers, and their diameters are ca. 1-10 μm. Fig. 1(b) shows the TiO$_2$ rods which are similar to that shown in Fig. 1(a), but there are a few of cracks on the rods are observed. This demonstrates that the titanium glycolate transformed into the TiO$_2$ without changing the rod-like morphology after calcining at 400 °C in air. As shown in Fig. 1(c), there are a lot of rods aggregate together and assemble into chrysanthemum-like structure. Fig. 1(d) shows the chrysanthemum-like microstructure which is composed of a number of rods. In addition, numerous rods connect with one root and form chsysanthemum-like morphology that can be seen in the inset of Fig. 1(d). It is clear that the diameters of the chrysanthemum are ca. 150-200 μm, and the rods are ca. 1-20 μm in diameter and ca. 100 μm in length which are larger than that of the titanium glycolate rods (Fig. 1(a)). The cross sections of the rods are irregular polygons as depicted in Fig. 1(d). The incomplete chrysanthemum-like structure clearly shows that the chrysanthemum is obtained by the titanium glycolate rods self-assembly growth. Fig. 1(e) and (f) show the similar results with the Fig. 1(c) and (d). It obviously indicates that the chrysanthemum-like structure was essentially preserved in the further calcined process.

Fig. 2 shows TEM images of $TiO_2$ rods, $TiO_2$/C chrysanthemums, and $TiO_2$ chrysanthemums, respectively. From the Fig. 2, it further reveals that the micrometer rods of all samples consist of many nanoparticles aggregated to nanoporous geometry. The $TiO_2$ rods are composed of many nanoparticles with diameters about 30 nm, as shown in Fig. 2(a). Fig. 2(b) shows the $TiO_2$/C chrysanthemums consist of numerous ultrafine nanoparticles with diameters less 10 nm. The $TiO_2$ chrysanthemums are composed of a lot of nanoparticles about 20 nm shown in Fig. 2(c). This demonstrates that the nanoparticles further grow and the chrysanthemum shape is still retained, when calcining the $TiO_2$/C chrysanthemums in air.

Fig. 3 shows the Raman spectra of the titanium glycolate rods, $TiO_2$ rods, $TiO_2$/C chrysanthemums and $TiO_2$ chrysanthemums, respectively. Fig. 3(a) shows the Raman spectrum of the titanium glycolate rods. All Raman peaks shown in Fig. 3(b) can be assigned as the $E_g$ (143 cm$^{-1}$), $B_{1g}$ (195 cm$^{-1}$), $A_{1g}$ (395 cm$^{-1}$), or $B_{1g}$ (515 cm$^{-1}$), and $E_g$ (638 cm$^{-1}$) modes of the anatase phase, respectively. As shown in Fig. 3(c), two strong peaks locate at 1356 and 1594 cm$^{-1}$ can be attributed to the carbon that derives from the carbonization of the organic compound, the other three peaks at 150 ($E_g$), 424 ($E_{1g}$) and 609 cm$^{-1}$ ($A_{1g}$) are attributed to the rutile phase. As depicted in Fig. 3(d), the peaks at 143 ($B_{1g}$), 237 (2 phonon process), 442 ($E_g$), and 610 cm$^{-1}$ ($A_{1g}$) are assigned to the rutile phase, respectively. The result showed that the as-prepared $TiO_2$ in this process is unstable due to the existence of carbon in the crystal. The carbon source was introduced from the alkoxide group and improves the transformation of crystallinity. These results are similar to Tseng Yao-Hsuan and co-worker's research [20]. It is clear that pure anatase rods and rutile chrysanthemum are obtained with heating the titanium glycolate rods at 400 $^{o}$C for 4 h in air and vacuum, respectively.

Fig. 4 shows the IR spectra of the same samples that were measured by Raman spectroscopy. The band at 3400 cm$^{-1}$ in Fig. 4(a) is assigned to the stretching of the O-H bond corresponding to physically absorbed water or EG, and the two sharp absorption bands at 2855 and 2925 cm$^{-1}$ are attributed to the alkyl-CH$_2$ symmetric and asymmetric stretching [13,14]. At approximately 1465 cm$^{-1}$ a signal corresponding to the bending of –CH$_2$ groups appears, and the bands at 1150-1350 cm$^{-1}$ are assigned to the scissoring of the C-H bonds of the -CH$_2$ groups [13,14]. In Fig 4(b), (c) and (d), the stretching of the C-H bonds in the TiO$_2$ rods is not observed, only the Ti-O bonds are remained. This indicates that the organic components were removed from the final products by the heattreatment.

It had been found that titanium alkoxide was added to EG and heated to 170 $^{o}$C for 2 h under rigorous stirring, the alkoxide was transformed into a chain-like, glycolate complex that subsequently crystallized into uniform nanowires or nanorods, then, titanium glycolate converts into TiO$_2$ without change its morphology [13-15]. However, to our surprise, we obtained the rutile TiO$_2$/C chrysanthemum-like structure by calcined the titanium glycolate rods at 400 $^{o}$C in vacuum. In this process, titanium glycolate rods were aggregated and self-assembled grown into the chrysanthemums with the dehydration and carbonization of organic groups of titanium glycolate rods in vacuum. As a result, the rods of chrysanthemums are larger than the titanium glycolate rods both in diameter and in length. Besides, the rods of the $TiO_2$/C chrysanthemums are composed of numerous ultrafine nanoparticles, which also demonstrated that the presence of carbon inhibits the growth of $TiO_2$ nanoparticles. We tentatively explain the growth mechanism of $TiO_2$/C chrysanthmums were caused by the much organic group of the titanium glycolate rods which undergoes the dehydration and carbonization process in vacuum rather than the dehydration and direct oxidation in air during the calcining process. Furthermore, the rutile $TiO_2$/C chrysanthemums transformed into pure rutile $TiO_2$ chrysanthemums without change in morphology by calcined in air. The nanoporous structures are attributed to the empty spaces that formed from the elimination of organic substance or carbon. In our case, the pure anatase $TiO_2$ rods and rutile $TiO_2$/C chrysanthemums were prepared by heating the titanium glycolate rods at 400 $^o$C in air and in vacuum, respectively. Then, the pure rutile $TiO_2$ chrysanthemums were obtained at lower temperature (400 $^o$C) by second heating the $TiO_2$/C chrysanthemums at 400 $^o$C in air. This phenomenon suggests that the carbon in the $TiO_2$/C chrysanthemums plays an important role in the phase transition which promotes the phase transformation toward rutile.

**4. Conclusions**

In summary, nanoporous anatase $TiO_2$ rods and rutile $TiO_2$ chrysanthemums were synthesized via an ethylene glycol-mediated synthesis route. The self-assembly growth process was found during the converting titanium glycolate rods into rutile $TiO_2$/C chrysanthemums in vacuum. The possible mechanism was proposed that the dehydration and carbonization of organic groups of titanium glycolate rods in vacuum result in the titanium glycolate rods aggregated and self-assembled grown into the rutile $TiO_2$/C chrysanthemum-like microstructure. The pure rutile $TiO_2$ chrysanthemums were obtained by calcining the $TiO_2$/C chrysanthemums at 400 $^o$C in air. It also indicates that the carbon plays an important role in the phase transition process. We believe that such a self-assembly process has large potential to prepare nanostructure metal oxides with novel morphologies.

**Acknowledgement**

This work was supported financially by (10674053, 10204010, 10574053), the Trans Century Training Programme Foundation for the Talents, RFDP (20020183050), the Cultivation Fund of the Key Scientific and Technical Innovation Project (2004-295) of MOE of China, the National Basic Research Program of China (2005 CB724400, 2001CB711201), and the Project for Scientific and Technical Development of Jilin province.

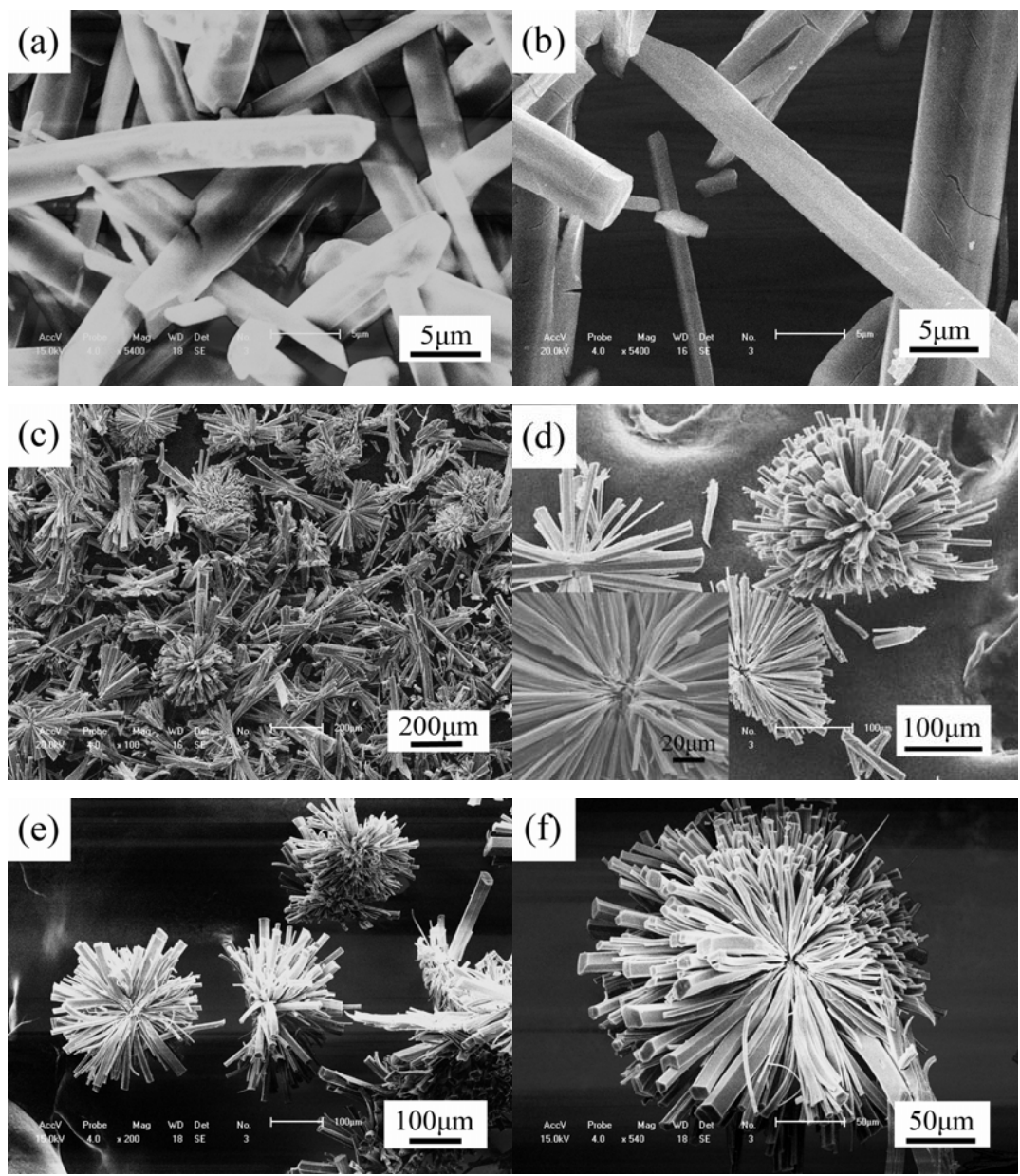

Fig. 1. SEM images of titanium glycolate rods (a), TiO$_2$ rods (b), self-assembly TiO$_2$/C chrysanthemums (c, d), and TiO$_2$ chrysanthemums (e, f). The inset of (d) is the high magnification image for the root of chrysanthemum.

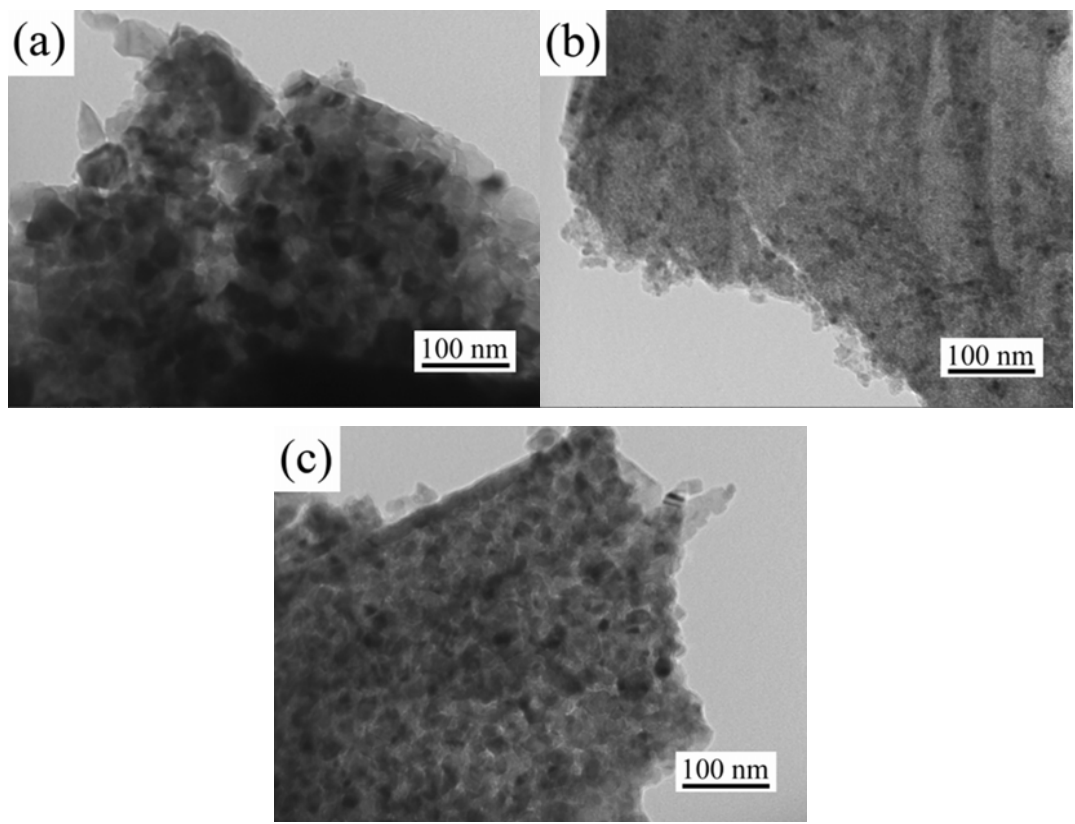

Fig. 2. TEM images of TiO$_2$ rods (a), TiO$_2$/C chrysanthemums (b) and TiO$_2$ chrysanthemums (c), respectively.

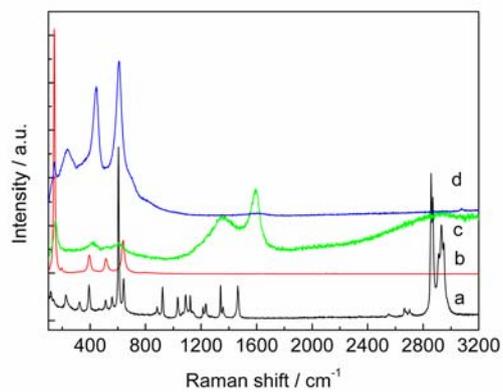

Fig. 3. Raman spectra of titanium glycolate rods (a), TiO$_2$ rods (b), self-assembled chrysanthemums (c), and TiO$_2$ chrysanthemums (d).

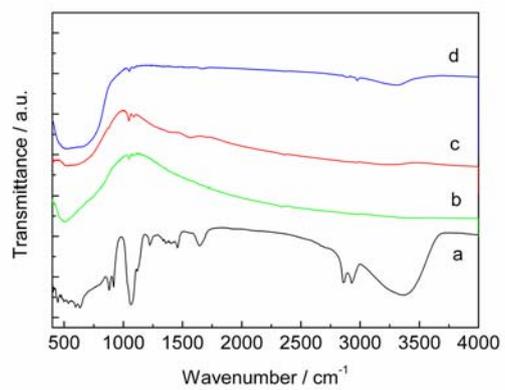

Fig. 4. IR spectra of titanium glycolate rods (a), TiO$_2$ rods (b), self-assembly TiO$_2$/C chrysanthemums (c), and TiO$_2$ chrysanthemums (d).